\documentclass[preprint,showpacs,preprintnumbers,amsmath,amssymb]{revtex4}

	\newif\ifpdf
	\ifx\pdfoutput\undefined
	\pdffalse 
	\else
	\pdfoutput=1 
	\pdftrue
	\fi

	\ifpdf
	\usepackage[pdftex]{graphicx}
	\else
	\usepackage{graphicx}
	\fi

\usepackage{graphicx}
\usepackage{dcolumn}
\usepackage{bm}

\begin{document}

	\ifpdf
	\DeclareGraphicsExtensions{.pdf, .jpg}
	\else
	\DeclareGraphicsExtensions{.eps, .jpg}
	\fi

\def\r{{\bf r}}

\preprint{}

\title{Aspects of diffusive-relaxation dynamics with a {\it non-uniform}, partially absorbing boundary in general porous media }
\author{Seungoh Ryu}
\author{David Linton Johnson}%
\affiliation{%
Schlumberger Doll Research\\
One Hampshire Street\\
Cambridge, MA 02139
}%

\date{\today}

\begin{abstract}
We consider the Helmholtz problem in the context of the evolution of uniform initial distribution of a physical attribute in general porous media subject to a partially absorbing boundary condition. Its spectral property as a reflection of  the boundary geometry has been widely exploited, such as in biological and geophysical applications. We consider the situation where the critical assumptions which enable such applications break down. Specifically, what are the consequences of an inhomogeneous absorption strength? 
By means of perturbation theory, exact theoretical results, and numerical simulations on random sphere packs, we identify the regions of parameter space in which such inhomogeneity is important and those in which it is not.
Our findings shed light on the issue that limits the mapping between the diffusion/relaxation spectrum and the underlying boundary geometry. 
\end{abstract}


\maketitle

The spatio-temporal evolution of the density of an attribute carried by diffusing agents, subject to a partially absorbing boundary, occurs in a variety of scientific context that ranges from NMR relaxometry in porous media\cite{Song2000,Grebenkov2007} and bio-medicine\cite{Lee2008a}, waves in a membrane\cite{Sapoval1991}, to migration of genes and cultural influences\cite{Fort1999a}. 
In its minimal form, the problem is formulated as the Helmholtz problem\cite{Arfken1970}, but in real systems, variations in the absorption strengths and local geometry complicate the matter. As a result, changes in part of its spectrum or {\em phantom} length scales, may appear. Empirical data from such systems, usually lacking the reference system with which to compare, beg the question: Is such a feature inherent in the nature of the underlying dynamics (with a clean boundary) or due to the haphazard elements on it? We are directly motivated by NMR relaxometry in porous media in which issues of such nature have been long standing. 
In this work, we consider the situation where the draining strength, $\rho(\r)$,  varies from point to point on the boundary surface and probe how it intertwines with the boundary geometry. Although some studies exist on aspects of variable $\rho(\r)$\cite{Wilkinson1991,Kansal2002,Valfouskaya2006,Arns2006}, their systematic investigation is still lacking.
In this letter, we develop a theoretical framework which incorporates both effects on an equal footing by treating the spatial fluctuations of $\rho(\r)$, namely $\delta \rho(\r)$, as a perturbative parameter and identify key spectral features observable through numerics or experiments.
The method is then applied to a simple problem, for which we obtain an exact solution for comparison. For realistic porous media and $\delta \rho(\r)$ variations, we perform numerical simulations to determine  bounds for the observable consequences. The result shows that the effect depends sensitively on the symmetry properties of both $\delta \rho(\r)$ and the eigenmodes of the boundary-value problem associated with the uniform $\rho$.

We start with a generic and widely studied problem: the evolution of a local density $\Psi(\r, t)$ of a scalar attribute carried by entities diffusing (with diffusivity $D$) inside a general pore space ($V_p$) defined by the pore-matrix interface $(\Sigma)$, which drains the attribute with a strength parametrized by the parameter $\rho_0 (> 0)$ upon contact with an entity. 
Defining the density operator as ${\bf J} \equiv - D(\r)\nabla$, and ${\cal H} \equiv \nabla \cdot {\bf J}$, the local continuity of $\Psi$ leads to the classic Helmholtz equation, and in seeking its solution in terms of the eigenmodes (i.e. $\{\phi_p^0 (\r) e^{- \lambda_p^0 t} \})$, one arrives at :
\begin{equation}
\label{eq:helmholtz}
{\cal H} \phi^0_p (\r) = \lambda_p^0 \phi_p^0 (\r)
\end{equation}
where each mode is subject to the Robin's condition 
$\hat{n}\cdot {\bf J} \phi_p (\r) = \rho_0 \phi_p^0(\r)$ on the boundary. We superscript  $\lambda_p^0$ and $\phi_p^0$ to indicate that it is for the case with a uniform $\rho_0$. 
From the properties of ${\cal H}$ and $\rho_0$, it follows that $\lambda_p^0$ are real, and $\ge 0$; orthonormal eigenmodes $\phi_p^0$ can be represented as real functions. 
The connection between the spectrum of ${\cal H}$  and the {\em boundary shape} (pore geometry) had been noted by Kac and others\cite{Kac1966,Gordon1992,Chapman1995} and some of its consequences were widely exploited in a variety scientific disciplines
\cite{Sapoval1999,Rocchesso2001,deGennes1982,Mitra1992}. In particular, the relaxation of polarized proton spins carried by fluid molecules in porous media such as rocks and biological samples is a pertinent example, as the relaxation is enhanced by the fluctuating dipole field near the interface.\cite{Kleinberg1996, Brownstein1979}
The evolution of the total attribute (magnetization in the case of NMR) ${\cal M} \equiv \int_{V_p} d\r \Psi(\r, t)$, if we assume it to be initially uniform, then follows ${\cal M}(t) = \sum_p e^{-\lambda_p t} |a_p^0|^2$ where $a^0_p = \int_{V_p} \phi_p^0 (\r) d\r$.  In the case of a simple, closed boundary with a single defining length scale (such as the radius $a$ in a spherical pore), this enhanced relaxation may be limited either by the strength of relaxation at the boundary or by the diffusivity $D$ and a control parameter $\kappa \equiv \rho_0 a / D$ emerges to separate the regimes which have distinct spectral properties (i.e. the weights $a_p$ and values of the slow modes $\lambda_p, p=0, 1, \ldots)$.  In the limit where $\kappa  \ll 1$, it was observed that $a_0 \sim 1$ for the slowest decay mode $(p = 0)$, and also that $\lambda_0^0  = \rho_0 S / V_p $ ($=\rho_0 3/ a$ for a sphere) directly proportional to the surface-to-volume ratio of the pore.  In the other limit, faster modes generally gain in weight, and $\lambda_0^0 \sim \lambda_\infty \equiv D \pi^2 / a^2$.
In many situations, the relationship $\lambda_0^0 \sim \rho / a$ is exploited to  map the observed spectral distribution $a_p, p = 0, 1, \ldots$ to a distribution of pore sizes. For this mapping to work, however, the pores should be isolated from each other (i.e. diffusive coupling \cite{Zielinski2002} is negligible), each satisfying the condition $\kappa \ll 1,$ and finally $\rho_0$ should be uniform. 
In an extended pore, the local variations in its geometry and the strength of the diffusive coupling make it no longer feasible to characterize the dynamics with a single length scale parameter $a$, and therefore the first two assumptions break down. Furthermore, in most real systems, $\rho$ assumes a spatial variation (we will call it {\em texture} from now on).  In addition to the fundamental issues as posed by Kac and others\cite{Kac1966,Gordon1992}, it is straightforward to show that, for a given distribution $a_p$ arising from a collection of pores distributed in sizes all satisfying $\kappa \ll 1$, one can construct a collection of identical pores with an appropriately chosen range of $\rho$ strengths assigned to each. 

In real life porous media, the pore space forms an extended, multiply connected manifold, with possible local variations in the connectivity of its constituents. It becomes necessary,  then to employ notions more than {\rm pore sizes and throats} for such systems, analogous to the way one progresses from atomic orbitals to the band theory and further on for disorder in solid state physics. In this approach, a few  properties of the eigenspectrum become key elements\cite{Ryu2001,Lisitza2002}. In the following, we seek to quantify how inhomogeneous $\rho(\r)$ affects the eigen-spectrum and its experimental manifestation.

Let us first generalize the Helmholtz problem Eq.\ref{eq:helmholtz} by substituting $\rho(\r) = \rho_0 + \delta \rho(\r)$ for $\rho_0$ with the requirement $\oint_\Sigma \delta\rho (\r) d\sigma = 0$. Under this  boundary condition, we pursue the new set of eigenmodes $\{ \phi_p \}$ with eigenvalues $\lambda_p$. 
Using the self-adjointedness of ${\cal H}$ and Green's theorem, we obtain the following relationship:\cite{Ryu2001,Ryu2008c}
\begin{equation}
\label{eq:lambda}
\lambda_p = \oint_\Sigma \rho (\r)  |\phi_p (\r)|^2 d\sigma   + \int_{V_p}  D(\r)  |\nabla \phi_p(\r) |^2 d\r
\end{equation}
which expresses all eigenvalues as a sum of surface integral and the diffusion-controlled volume integral, analogous to the energy of a particle in a potential well given in terms of the potential and the kinetic part. The character of each eigenmode can be understood in terms of the competition between these two components. A general observation can be made for fast modes that the second component dominates and $\lambda_p$ for $p > 0$ becomes progressively insensitive to $\rho_0$ and $\delta \rho$.
We define $\kappa_p  =  \frac{\rho_0 \ell_p }{D}$ by introducing a length scale parameter for each mode 
$\ell_p \equiv  \frac{\int d\r (\nabla \phi_p)^2}{\oint d\sigma (\nabla \phi_p)^2}$. Applied to the $p=0$ mode, the criterion $\kappa_0 \ll \, {\rm or}\,  \gg 1$  generalizes the earlier observations made for simple closed pore geometry\cite{Brownstein1979} and is reminiscent of the $\Lambda$ parameter for the electrical conductivity of pore filling fluid\cite{Johnson1986}. The spectral weight for the slowest mode $a_0$ is significantly weakened for $\kappa_0 \gg 1$, which leads to ${\cal M}(t)$ with a {\em multi-exponential} characteristics that had invited the {\em potentially misleading} interpretation based on isolated pore size distributions. Instead, we derive a relationship that shows that this weight is directly related to the spatial fluctuation of the slowest eigenmode:
\begin{equation}
\label{eq:ap}
|a_0|^2 = 1 -  V_p \Big( \int_{V_p} d\r |\phi_0(\r)|^2  - | \int_{V_p} d\r \phi_0(\r) |^2 \Big).
\end{equation}
Note that these rigorous relationships (Eq.\ref{eq:lambda}, \ref{eq:ap}) apply to general boundary shape, and both uniform and inhomogeneous $\rho$. 
It is also straightforward to prove that slope of $\log {\cal M}(t)$ at early times should remain robust against the fluctuatons $\delta\rho(\r)$, 
$- \lim_{t\rightarrow 0}  \frac{d}{dt} {\log \cal M}(t) \rightarrow  \rho_0 \frac{S}{{\cal V}}$
but the range over which this is valid could be severely limited depending on the strength of $|\delta \rho|$.
Many authors had considered the 
so-called mean lifetime $\tau = \sum_p a_p^2 /\lambda_p$ \cite{Wilkinson1991,Kansal2002,Grebenkov2007}, 
for which we obtain 
$\tau = \tau_0 - \frac{1}{V_p} \oint_\Sigma d\sigma u_0^0 (\r) \delta \rho( \r) u_0(\r)
$ where $u_0 \equiv \lim_{s=0} \int_0^\infty \Psi(\r, t) e^{-t s} dt$ with $\Psi(\r, t)$ being the local density under $\rho(\r)$, and similarly with $u_0^0$ and $\Psi_0$ under the uniform $\rho_0$. 

\begin{figure}[p]
\includegraphics[width=6in]{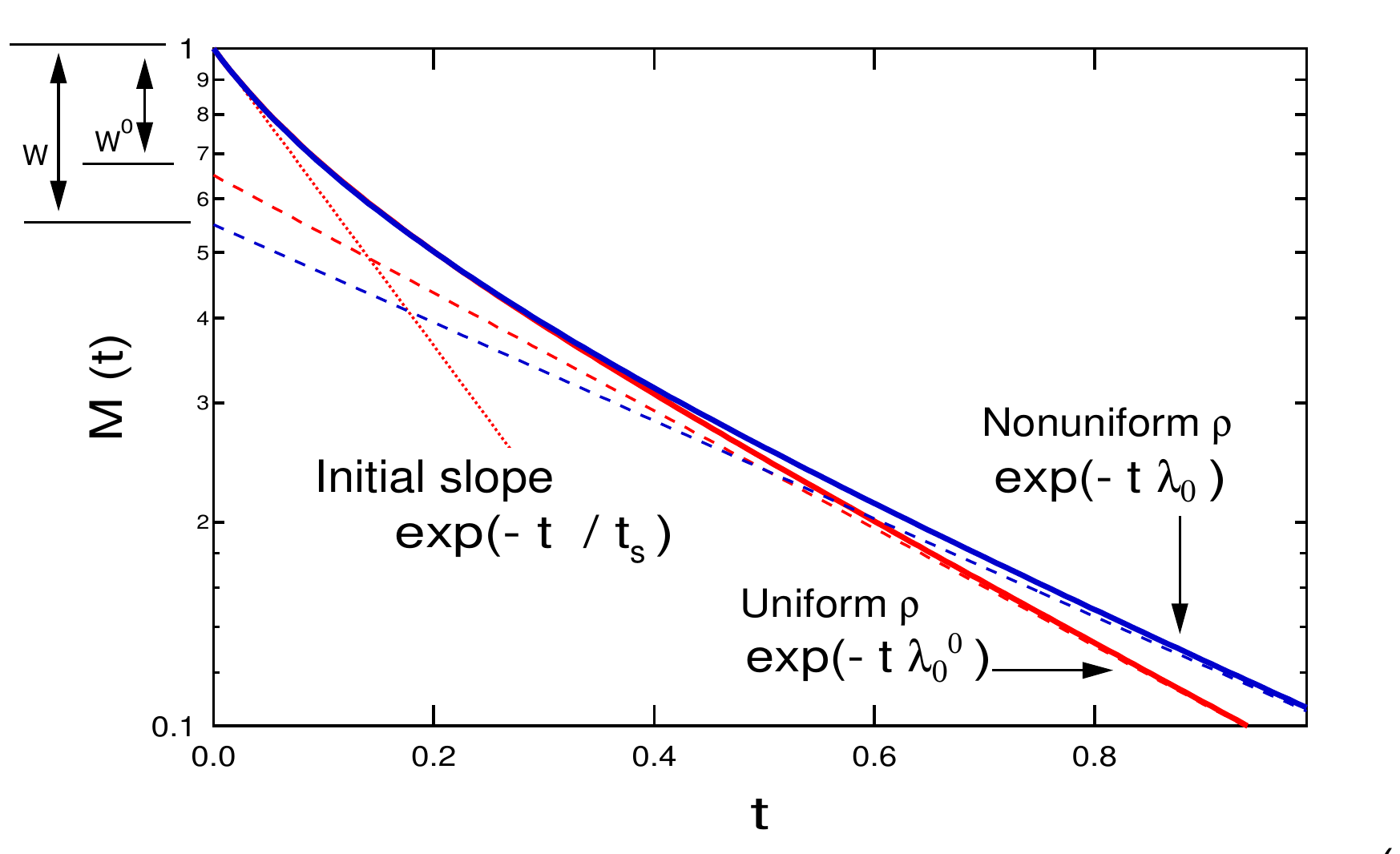}
\caption{\label{fig:schematics}Schematics for the difference between the population evolution with uniform $\rho_0$ and an inhomogeneous $\rho(\r)$ with the finial slopes given by $\lambda_0^0$ and $\lambda_0$ as indicated by the broken curves. The accompanying change in the spectral weight distribution is reflected in the difference $W - W^0$ on the y-axis where  $W \equiv 1 - (a_0)^2$ and 
 $W^0 \equiv 1 - (a^0_0)^2$.  }
\end{figure}

The eigenvalue $\lambda_0$ and the spectral weight $|a_0|^2$ of the slowest mode are the most accessible indicator for the change in the boundary condition $\rho_0 \rightarrow \rho_0 + \delta \rho(\r)$. In the following, we therefore focus on the fractional shift in the slowest eigenmode $\frac{\delta \lambda_0}{\lambda^0_0} \equiv \frac{\lambda^0_0 - \lambda_0}{\lambda^0_0}$ which determines the long-time slope of $\log {\cal M}(t)$ vs. $t$. Figure \ref{fig:schematics} summarises schematically these general observations.
We first derive a perturbative solution for $\frac{\delta \lambda_0}{\lambda^0_0}$ for an arbitrary pore geometry and $\rho(\r)$ texture, and compare the result with an exact solution. 
Assuming that the complete eigenmodes $\{\phi_p^0\}$ with $\lambda_p^0$'s are worked out already for the uniform $\rho_0$, we put the eigenmodes for the new boundary condition with $\rho(\r)$ as 
\begin{equation}
\label{eq:definenewsol}
\phi_p (\r) =
c_p \Big(  \phi_p^0 (\r) + \sum_{q\ne p} a_{pq} \phi_q^0 (\r) \Big) + Q_p (\r)
\end{equation}
where $c_p$ is the normalization constant. We introduce the auxiliary function $ Q_p (\r)  \equiv  \phi_p (\r) - \int d\r' {\cal P}(\r,\r') \phi_p(\r') $ defined via the projection operator ${\cal P}(\r,\r') = \sum_p \phi_p^0 (\r') \phi_p^0 (\r)$ onto the Hilbert space spanned by the eigenmodes $\{\phi_p^0\}$ of the uniform case.
Note that formal inclusion of $Q_p(\r)$ is necessary at this point to satisfy the new boundary condition as $\phi_p$, if it were to be spanned by $\{\phi_p^0\}$ alone, would satisfy the uniform $\rho_0$ condition.
Using the orthonormality of the complete sets $\{ \phi_p \}$ and $\{ \phi_p^0\}$ respectively, it is straightforward to obtain a recursive equation for $a_{pq}$:
\begin{eqnarray}
\label{eq:gamma1}
a_{pq }&=&   \frac{(1-\delta_{pq}) }{\lambda_p - \lambda_q^0} \Big\{ \sum_r a_{pr} \delta\rho_{qr} + \delta \tilde\rho_{qp}  \Big\}  \frac{S}{V_p} + \delta_{pq}
\end{eqnarray}
where we define overlap integrals $\delta\rho_{qr} \frac{S}{V_p} \equiv \oint_\Sigma d\sigma \phi_q^0  \delta \rho  \phi_r^0$
and
$\delta \tilde \rho_{qp} \frac{S}{V_p}  \equiv \oint d\sigma_\Sigma \phi_q^0  \delta \rho  Q_p$.
Via iterative substitutions, we obtain the desired result in a systematic power expansion in $\delta \rho$:
\begin{equation}
\label{eq:eqforlambda}
\lambda_p 
=   \lambda_p^0 + \frac{S}{V_p} \Big( \delta\rho_{pp}   -  \sum_{q\ne p} \frac{\delta \rho_{pq}  \delta\rho_{qp}}
{ \lambda_q^0 - \lambda^0_p}  \frac{S}{V_p} +  \delta \tilde\rho_{pp} \Big) + {\cal O}(\delta\rho^3).
 \end{equation}
Defining $ f_p(\r)  = \frac{1}{c_p} \sum_{q}  \phi_q^0 (\r) \oint_\Sigma d\sigma  \phi_q^0  \delta \rho \phi^0_p$, a representation of $\delta \rho(\r)$ projected onto the Hilbert space spanned by $\{\phi_p^0\}$, we find that $Q_p$ should satisfy the inhomogeneous equation: 
$
 ({\cal H} - \lambda_p) Q_p (\r)  =  f_p(\r) 
$ subject to the condition
$ \rho_0 Q_p (\r)   -  \hat{n} (\r) \cdot {\bf J}  Q_p (\r)  = - \frac{1}{c_p} \delta \rho(\r) \phi^0_p (\r)
$ on the boundary.
As $Q_p = [ 1 - {\cal P}] \phi_p$, its perturbative solution shows that it can be viewed as a superposition of waves with wavelength $\sqrt{ D/ \lambda_p}$ emanating from a surface localized source $[ 1 - {\cal P}] \delta \rho(\r)$.  For $p=0$, our main focus, the effective source is averaged over a diffusion length $\sqrt{D/\lambda_0}$, as indicated by the presence of the $\lambda_p Q_p (\r)$ term in its governing equation. This leads to $\delta \tilde{\rho}_{00} \sim 0$ as we find in the perturbative solution of the spherical pore\cite{Ryu2008c} and also from exact evaluations\cite{Johnson2008}. 
Without any assumptions on the pore geometry or the $\rho(\r)$ texture, we make some general observations for each contribution to $\lambda_p$.
Note that the first order term, $\delta\rho_{00}$, depends sensitively on the symmetry and the profile of the mode, $(\phi_0^0)^2$, along the boundary in relation to $\delta \rho$. While it vanishes for the simple situations where $\phi_0^0$ is uniform along the boundary, it may not do so when there exists significant variation of $\phi_0$ as when the complex pore geometry dictates. For empirical ${\cal M}(t)$ with a multi-exponential characteristics, as often observed in geophysical applications\cite{Kenyon1992},  Eq.\ref{eq:ap} suggests  that one cannot safely assume $\phi_0^0(\r)$ is uniform along the contours of the boundary. The chance for a sizeable first order contribution is further enhanced when the texture $\delta\rho(\r)$ varies commensurate with the former. 

\begin{figure}[p]
\includegraphics[width=6in]{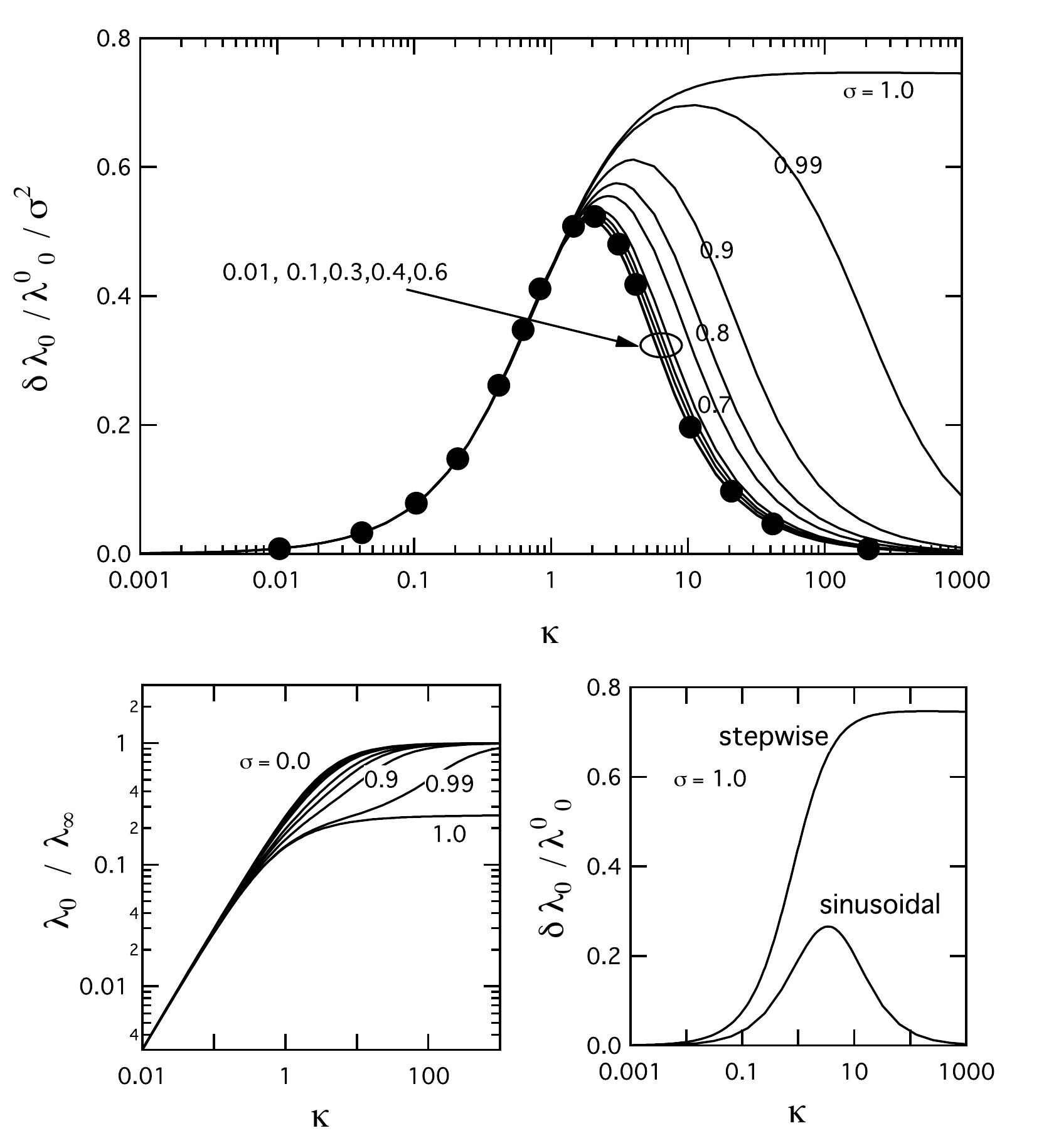}
\caption{\label{fig:exact} Top: Results for a sphere of radius $a$ and varying $\kappa = \rho_0 a / D$ values with the hemispherical $\rho$ texture.  The solid lines show $\delta \lambda_0/\lambda_0^0$ for $\sigma = 0.01 \sim  1.0$ as obtained from the exact matrix formulation described in the text. The filled points represent the second order perturbation result. Bottom left panel shows the $\lambda_0/\lambda_\infty$  ( $\lambda_\infty = D \frac{\pi^2}{a^2}$ ) for varying $\sigma$'s as above.  Bottom-right panel compares $\delta \lambda_0 / \lambda_0^0$ with $\sigma = 1.0$ for the hemispherical and the sinusoidal textures of $\rho(\r) =\rho_0 (1 + \sigma \cos (\theta))$.}
\end{figure}

Now we turn to a spherical pore of radius $a$,  and seek exact solutions for both uniform $\rho_0$ and $\rho(\r)$ of the form (with $\sigma \le 1$)
$\rho(\r) = \rho_0 ( 1  + \sigma f(\theta) )$
where we consider the cases of a stepwise texture ($f(\theta < \pi/2) = -1, f(\theta \ge \pi/2) = 1)$) and a sinusoidal ($f(\theta) = \cos(\theta)$).
We look for the eigenmodes\cite{Johnson2008} in the form of 
$\phi_k(\r) = \sum_{L=0}^\infty s_{k,L} j_L(k  r) Y_L^0(\Omega)
$
where $j_L(x)$ is the spherical Bessel function, $Y_L^0$'s are the spherical harmonic functions with $M=0$ (due to the azimuthal symmetry). $k$ represents an infinite set of numbers that allow for a non-trivial solution for the coefficients ${\bf s}_{k}$ that facilitate the boundary condition be met:
\begin{eqnarray}
\label{eq:sphericalbc}
2 \kappa \sum_L \triangle_{L,L'} j_L(k a) s_{k,L} - ( j_{L'}(k a) &+& k a  (j_{L'+1}(k a) 
\nonumber \\   - j_{L'-1}(k a) ) s_{k,L'} &=& 0 
\end{eqnarray}
where $\triangle_{L,L'} =  \int  d\Omega f(\theta) Y_{L'}^0 Y_{L}^0$. 
Viewed as a homogeneous matrix equation ${\cal K} \cdot {\bf s}_k = 0$, the eigenvalues are found from the condition that det $ [{\cal K} ]= 0$. We solve this by truncating the matrix to a finite though large size and searching numerically for the root. The fractional difference in the lowest eigenvalue ($\lambda_0 = D k_{min}^2$) between the uniform and non-uniform cases are shown in Figure \ref{fig:exact}. First panel shows the {\em stepwise} texture with $\sigma$ ranging from $ 0.01$ to $1.0$ as indicated. The results from both the exact solution (solid lines) and the second order perturbation (points) agree very well for $\kappa < 2$ for all values of $\sigma$,  while for $\kappa > 2$, the agreement deteriorates progressively as $\sigma$ grows beyond 0.5.  $\lambda_0$ itself is shown in the second panel.  Anomaly occurs   in the $\kappa \gg 1$ limit with $\sigma = 1$ for which  $\rho(\r)$ vanishes on half of the hemisphere. In this special case, the limit  $\kappa \gg 1$ acquires a new {\em diffusion controlled} time scale (i.e. $\sim 1/\lambda_0$ is quadrupled from $\lambda_\infty = \frac{a^2}{D\pi^2}$ to $\lambda_\infty / 4$, as the slowest mode is now controlled by diffusion from the $\rho = 0$ zone to the other end $\rho \rightarrow \infty$ over the distance of $2 a$. This  crossover is missing for the sinusoidal texture with $\sigma=1$ (in the third  panel)  for which only a nodal point ($\theta = \pi$) exists on which $\rho(\r)$ vanishes. This contrasting behavior is also verified in numerical simulations. 
Using the perturbative approach, it is now straightforward to incorporate more complicated $\rho(\r)$ textures for small $\sigma$,  using Figure \ref{fig:exact} as a guide for its validity.

\begin{figure}[p]
\includegraphics[width=6in]{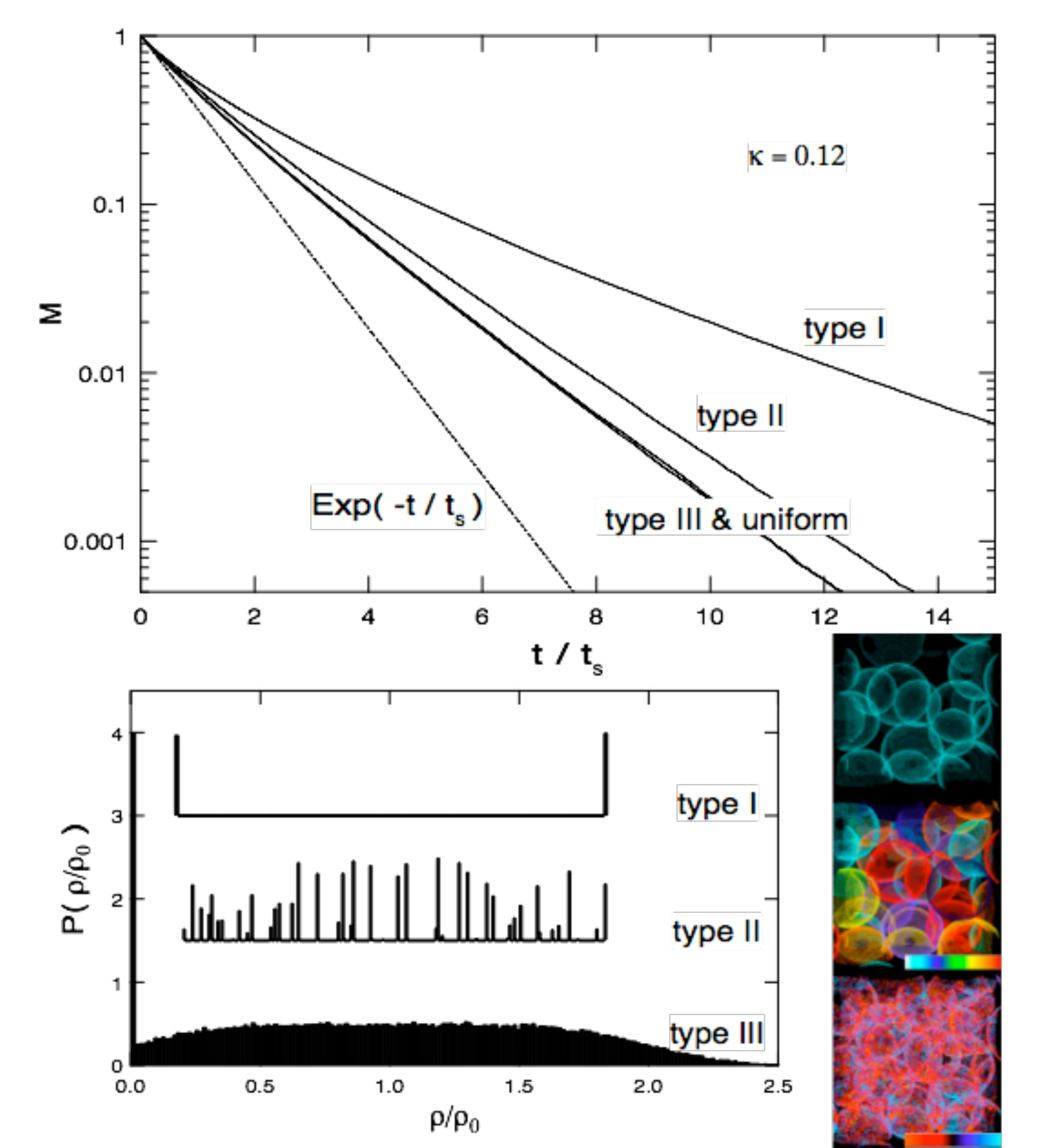}
\caption{\label{fig:finney} Top: Results for a Finney pack with three differenct  $\rho$ textures (for all $\rho_0 V_p/S / D = 0.12) $ as described in the text.  Also shown is the result with the matching uniform $\rho_0$ for comparison. The broken line is an exponential function with the initial slope as expected with $1/t_s = \rho_0 S/ V_p.$ }
\end{figure}

Next, we consider the random glass bead pack as an example of realistic porous media for which well controlled experiment and simulations could be carried out. Figure \ref{fig:finney} shows the results from randomwalk simulations\cite{Ryu2008b} based on the Finney pack\cite{Finney1968} in which we realize three different $\rho$ textures that clearly violate the conditions necessary for the perturbative approach. Type I shows the strongest deviation from the uniform case (and is analogous to the case of the hemispherical $\rho$ of Fig.\ref{fig:exact} with $\sigma \sim 1$) as we randomly assign a value of $0.16$ or $1.84 \times \rho_0$ to each grain with equal probability. In this case, $\delta\lambda_0/\lambda_0^0/\sigma^2\sim 1.22$ is significantly larger than in the closed sphere even though $\kappa \sim 0.12$.  This is likely due to the existence of wider spatial separations between the two $\rho$ values, as expected in the pore morphology of a random packing. Type II draws randomly from a distribution of $\rho$ values. Type III uses a texture generated using a correlated random noise\cite{Ryu2008b}. In this case, values of $\delta\rho(\r)$ are correlated over just a fraction of the bead radius, separating the correlations of boundary shape variation from that of $\delta\rho(\r)$. Note that even though such $\delta \rho(\r)$ has a wider distribution (Bottom panels show the histogram and graphical rendition of each), the diffusive smearing greatly reduces its effect, and we obtain a result virtually indistinguishable from the uniform case. Similar observation had been made numerically by Valfouskaya {\em et al}\cite{Valfouskaya2006}. It is beyond the scope of this letter to delve into the relevance of these textures for real systems, but they  cover a wide range of plausible patterns present in natural media such as rocks and certainly realizable  in bead packs\cite{Godefroy2001a} and other artificial structures.\cite{Lee2008a} Quantitative application of our perturbative solution to realistic 3D systems will require the profile of the relevant eigenmodes on the boundary, which may rarely be available analytically, but through numerical evaluations of $\phi^0_p$'s.\cite{Hoop2007,Ryu2008b}
\begin{acknowledgments}
We wish to acknowledge professor Adrianus T. de Hoop for useful discussions. 
\end{acknowledgments}

\bibliography{strass}

\begin{thebibliography}{30}
\expandafter\ifx\csname natexlab\endcsname\relax\def\natexlab#1{#1}\fi
\expandafter\ifx\csname bibnamefont\endcsname\relax
  \def\bibnamefont#1{#1}\fi
\expandafter\ifx\csname bibfnamefont\endcsname\relax
  \def\bibfnamefont#1{#1}\fi
\expandafter\ifx\csname citenamefont\endcsname\relax
  \def\citenamefont#1{#1}\fi
\expandafter\ifx\csname url\endcsname\relax
  \def\url#1{\texttt{#1}}\fi
\expandafter\ifx\csname urlprefix\endcsname\relax\def\urlprefix{URL }\fi
\providecommand{\bibinfo}[2]{#2}
\providecommand{\eprint}[2][]{\url{#2}}

\bibitem[{\citenamefont{Grebenkov}(2007)}]{Grebenkov2007}
\bibinfo{author}{\bibfnamefont{D.~S.} \bibnamefont{Grebenkov}},
  \bibinfo{journal}{Rev. Mod. Phys.} \textbf{\bibinfo{volume}{79}},
  \bibinfo{pages}{1077} (\bibinfo{year}{2007}).

\bibitem[{\citenamefont{Song et~al.}(2000)\citenamefont{Song, Ryu, and
  Sen}}]{Song2000}
\bibinfo{author}{\bibfnamefont{Y.-Q.} \bibnamefont{Song}},
  \bibinfo{author}{\bibfnamefont{S.}~\bibnamefont{Ryu}}, \bibnamefont{and}
  \bibinfo{author}{\bibfnamefont{P.~N.} \bibnamefont{Sen}},
  \bibinfo{journal}{Nature} \textbf{\bibinfo{volume}{406}},
  \bibinfo{pages}{178} (\bibinfo{year}{2000}).

\bibitem[{\citenamefont{Lee et~al.}(2008)\citenamefont{Lee, Sun, Ham, and
  Weissleder}}]{Lee2008a}
\bibinfo{author}{\bibfnamefont{H.}~\bibnamefont{Lee}},
  \bibinfo{author}{\bibfnamefont{E.}~\bibnamefont{Sun}},
  \bibinfo{author}{\bibfnamefont{D.}~\bibnamefont{Ham}}, \bibnamefont{and}
  \bibinfo{author}{\bibfnamefont{R.}~\bibnamefont{Weissleder}},
  \bibinfo{journal}{Nature Medicine} \textbf{\bibinfo{volume}{14}},
  \bibinfo{pages}{869} (\bibinfo{year}{2008}).

\bibitem[{\citenamefont{Sapoval et~al.}(1991)\citenamefont{Sapoval, Gobron, and
  Margolina}}]{Sapoval1991}
\bibinfo{author}{\bibfnamefont{B.}~\bibnamefont{Sapoval}},
  \bibinfo{author}{\bibfnamefont{T.}~\bibnamefont{Gobron}}, \bibnamefont{and}
  \bibinfo{author}{\bibfnamefont{A.}~\bibnamefont{Margolina}},
  \bibinfo{journal}{Phys. Rev. Lett.} \textbf{\bibinfo{volume}{67}},
  \bibinfo{pages}{2974} (\bibinfo{year}{1991}).

\bibitem[{\citenamefont{Fort and M{\'e}ndez}(1999)}]{Fort1999a}
\bibinfo{author}{\bibfnamefont{J.}~\bibnamefont{Fort}} \bibnamefont{and}
  \bibinfo{author}{\bibfnamefont{V.}~\bibnamefont{M{\'e}ndez}},
  \bibinfo{journal}{Phys. Rev. Lett.} \textbf{\bibinfo{volume}{82}},
  \bibinfo{pages}{867} (\bibinfo{year}{1999}).

\bibitem[{\citenamefont{Arfken}(1970)}]{Arfken1970}
\bibinfo{author}{\bibfnamefont{G.}~\bibnamefont{Arfken}},
  \emph{\bibinfo{title}{Mathematical Methods for Physicists}}
  (\bibinfo{publisher}{Academic Press}, \bibinfo{address}{New York},
  \bibinfo{year}{1970}).

\bibitem[{\citenamefont{Wilkinson et~al.}(1991)\citenamefont{Wilkinson,
  Johnson, and Schwartz}}]{Wilkinson1991}
\bibinfo{author}{\bibfnamefont{D.~J.} \bibnamefont{Wilkinson}},
  \bibinfo{author}{\bibfnamefont{D.~L.} \bibnamefont{Johnson}},
  \bibnamefont{and} \bibinfo{author}{\bibfnamefont{L.~M.}
  \bibnamefont{Schwartz}}, \bibinfo{journal}{Phys. Rev. B}
  \textbf{\bibinfo{volume}{44}}, \bibinfo{pages}{4960} (\bibinfo{year}{1991}).

\bibitem[{\citenamefont{Kansal and Torquato}(2002)}]{Kansal2002}
\bibinfo{author}{\bibfnamefont{A.~R.} \bibnamefont{Kansal}} \bibnamefont{and}
  \bibinfo{author}{\bibfnamefont{S.}~\bibnamefont{Torquato}},
  \bibinfo{journal}{J. Chem. Phys.} \textbf{\bibinfo{volume}{116}},
  \bibinfo{pages}{10589} (\bibinfo{year}{2002}).

\bibitem[{\citenamefont{Valfouskaya et~al.}(2006)\citenamefont{Valfouskaya,
  Adler, Thovert, and Fleury}}]{Valfouskaya2006}
\bibinfo{author}{\bibfnamefont{A.}~\bibnamefont{Valfouskaya}},
  \bibinfo{author}{\bibfnamefont{P.~M.} \bibnamefont{Adler}},
  \bibinfo{author}{\bibfnamefont{J.~F.} \bibnamefont{Thovert}},
  \bibnamefont{and} \bibinfo{author}{\bibfnamefont{M.}~\bibnamefont{Fleury}},
  \bibinfo{journal}{J. Coll. Interf. Sci.} \textbf{\bibinfo{volume}{295}},
  \bibinfo{pages}{188} (\bibinfo{year}{2006}).

\bibitem[{\citenamefont{Arns et~al.}(2006)\citenamefont{Arns, Sheppard,
  Saadatfar, and Knackstedt}}]{Arns2006}
\bibinfo{author}{\bibfnamefont{C.~H.} \bibnamefont{Arns}},
  \bibinfo{author}{\bibfnamefont{A.~P.} \bibnamefont{Sheppard}},
  \bibinfo{author}{\bibfnamefont{M.}~\bibnamefont{Saadatfar}},
  \bibnamefont{and} \bibinfo{author}{\bibfnamefont{M.~A.}
  \bibnamefont{Knackstedt}}, \bibinfo{journal}{SPWLA 47th Annual Logging
  Symposium} p. \bibinfo{pages}{498610GG} (\bibinfo{year}{2006}).

\bibitem[{\citenamefont{Kac}(1966)}]{Kac1966}
\bibinfo{author}{\bibfnamefont{M.}~\bibnamefont{Kac}}, \bibinfo{journal}{Am.
  Math. Mon.} \textbf{\bibinfo{volume}{73}}, \bibinfo{pages}{1}
  (\bibinfo{year}{1966}).

\bibitem[{\citenamefont{Gordon et~al.}(1992)\citenamefont{Gordon, Webb, and
  Wolpert}}]{Gordon1992}
\bibinfo{author}{\bibfnamefont{C.}~\bibnamefont{Gordon}},
  \bibinfo{author}{\bibfnamefont{D.~L.} \bibnamefont{Webb}}, \bibnamefont{and}
  \bibinfo{author}{\bibfnamefont{S.}~\bibnamefont{Wolpert}},
  \bibinfo{journal}{Bull. Am. Math. Soc.} \textbf{\bibinfo{volume}{27}},
  \bibinfo{pages}{134} (\bibinfo{year}{1992}).

\bibitem[{\citenamefont{Chapman}(1995)}]{Chapman1995}
\bibinfo{author}{\bibfnamefont{S.~J.} \bibnamefont{Chapman}},
  \bibinfo{journal}{Am. Math. Mon.} \textbf{\bibinfo{volume}{102}},
  \bibinfo{pages}{124} (\bibinfo{year}{1995}).

\bibitem[{\citenamefont{Sapoval et~al.}(1999)\citenamefont{Sapoval, Filoche,
  Karamanos, and Brizzi}}]{Sapoval1999}
\bibinfo{author}{\bibfnamefont{B.}~\bibnamefont{Sapoval}},
  \bibinfo{author}{\bibfnamefont{M.}~\bibnamefont{Filoche}},
  \bibinfo{author}{\bibfnamefont{K.}~\bibnamefont{Karamanos}},
  \bibnamefont{and} \bibinfo{author}{\bibfnamefont{R.}~\bibnamefont{Brizzi}},
  \bibinfo{journal}{Eur. Phys. J.} \textbf{\bibinfo{volume}{B 9}},
  \bibinfo{pages}{739} (\bibinfo{year}{1999}).

\bibitem[{\citenamefont{Rocchesso}(2001)}]{Rocchesso2001}
\bibinfo{author}{\bibfnamefont{D.}~\bibnamefont{Rocchesso}},
  \bibinfo{journal}{Proc. of the 2001 Int. Conf. on Auditory Display, Espoo,
  Finland}  (\bibinfo{year}{2001}).

\bibitem[{\citenamefont{de~Gennes}(1982)}]{deGennes1982}
\bibinfo{author}{\bibfnamefont{P.~G.} \bibnamefont{de~Gennes}},
  \bibinfo{journal}{C. R. Acad. Sc. Paris} \textbf{\bibinfo{volume}{295}},
  \bibinfo{pages}{1061} (\bibinfo{year}{1982}).

\bibitem[{\citenamefont{Mitra and Sen}(1992)}]{Mitra1992}
\bibinfo{author}{\bibfnamefont{P.~P.} \bibnamefont{Mitra}} \bibnamefont{and}
  \bibinfo{author}{\bibfnamefont{P.~N.} \bibnamefont{Sen}},
  \bibinfo{journal}{Phys. Rev. B} \textbf{\bibinfo{volume}{45}},
  \bibinfo{pages}{143} (\bibinfo{year}{1992}).

\bibitem[{\citenamefont{Kleinberg}(1996)}]{Kleinberg1996}
\bibinfo{author}{\bibfnamefont{R.~L.} \bibnamefont{Kleinberg}}, in
  \emph{\bibinfo{booktitle}{Encyclopedia of Nuclear Magnetic Resonance}},
  edited by \bibinfo{editor}{\bibfnamefont{D.~M.} \bibnamefont{Grant}}
  \bibnamefont{and} \bibinfo{editor}{\bibfnamefont{R.~K.} \bibnamefont{Harris}}
  (\bibinfo{publisher}{John Wiley}, \bibinfo{address}{Chichester},
  \bibinfo{year}{1996}).

\bibitem[{\citenamefont{Brownstein and Tarr}(1979)}]{Brownstein1979}
\bibinfo{author}{\bibfnamefont{K.~R.} \bibnamefont{Brownstein}}
  \bibnamefont{and} \bibinfo{author}{\bibfnamefont{C.~E.} \bibnamefont{Tarr}},
  \bibinfo{journal}{Phys. Rev. A} \textbf{\bibinfo{volume}{19}},
  \bibinfo{pages}{2446} (\bibinfo{year}{1979}).

\bibitem[{\citenamefont{Zielinski et~al.}(2002)\citenamefont{Zielinski, Song,
  Ryu, and Sen}}]{Zielinski2002}
\bibinfo{author}{\bibfnamefont{L.~J.} \bibnamefont{Zielinski}},
  \bibinfo{author}{\bibfnamefont{Y.-Q.} \bibnamefont{Song}},
  \bibinfo{author}{\bibfnamefont{S.}~\bibnamefont{Ryu}}, \bibnamefont{and}
  \bibinfo{author}{\bibfnamefont{P.~N.} \bibnamefont{Sen}},
  \bibinfo{journal}{J. of Chem. Phys.} \textbf{\bibinfo{volume}{117}},
  \bibinfo{pages}{5361} (\bibinfo{year}{2002}).

\bibitem[{\citenamefont{Ryu}(2001)}]{Ryu2001}
\bibinfo{author}{\bibfnamefont{S.}~\bibnamefont{Ryu}}, \bibinfo{journal}{Mag.
  Res. Imag.} \textbf{\bibinfo{volume}{19}}, \bibinfo{pages}{411}
  (\bibinfo{year}{2001}).

\bibitem[{\citenamefont{Lisitza and Song}(2002)}]{Lisitza2002}
\bibinfo{author}{\bibfnamefont{N.~V.} \bibnamefont{Lisitza}} \bibnamefont{and}
  \bibinfo{author}{\bibfnamefont{Y.-Q.} \bibnamefont{Song}},
  \bibinfo{journal}{Phys. Rev. B} \textbf{\bibinfo{volume}{65}},
  \bibinfo{pages}{172406} (\bibinfo{year}{2002}).

\bibitem[{\citenamefont{Ryu}(2009)}]{Ryu2008c}
\bibinfo{author}{\bibfnamefont{S.}~\bibnamefont{Ryu}},
  \bibinfo{journal}{submitted to Phys. Rev. E, also
  http://arxiv.org/abs/0903.1655}  (\bibinfo{year}{2009}).

\bibitem[{\citenamefont{Johnson et~al.}(1986)\citenamefont{Johnson, Koplik, and
  Schwartz}}]{Johnson1986}
\bibinfo{author}{\bibfnamefont{D.~L.} \bibnamefont{Johnson}},
  \bibinfo{author}{\bibfnamefont{J.}~\bibnamefont{Koplik}}, \bibnamefont{and}
  \bibinfo{author}{\bibfnamefont{L.~M.} \bibnamefont{Schwartz}},
  \bibinfo{journal}{Phys. Rev. Lett.} \textbf{\bibinfo{volume}{57}},
  \bibinfo{pages}{2564} (\bibinfo{year}{1986}).

\bibitem[{\citenamefont{Johnson and Ryu}(2008)}]{Johnson2008}
\bibinfo{author}{\bibfnamefont{D.~L.} \bibnamefont{Johnson}} \bibnamefont{and}
  \bibinfo{author}{\bibfnamefont{S.}~\bibnamefont{Ryu}},
  \bibinfo{journal}{unpublished}  (\bibinfo{year}{2008}).

\bibitem[{\citenamefont{Kenyon}(1992)}]{Kenyon1992}
\bibinfo{author}{\bibfnamefont{W.~E.} \bibnamefont{Kenyon}},
  \bibinfo{journal}{Nucl. Geophys.} \textbf{\bibinfo{volume}{6}},
  \bibinfo{pages}{153} (\bibinfo{year}{1992}).

\bibitem[{\citenamefont{Ryu}(2008)}]{Ryu2008b}
\bibinfo{author}{\bibfnamefont{S.}~\bibnamefont{Ryu}}, \bibinfo{journal}{SPWLA
  Proceedings of the 49th Annual Logging Symposium, SPWLA} p.
  \bibinfo{pages}{737008 BB} (\bibinfo{year}{2008}).

\bibitem[{\citenamefont{Finney}(1968)}]{Finney1968}
\bibinfo{author}{\bibfnamefont{J.~L.} \bibnamefont{Finney}}, Ph.D. thesis,
  \bibinfo{school}{University of London}, \bibinfo{address}{London}
  (\bibinfo{year}{1968}).

\bibitem[{\citenamefont{Godefroy et~al.}(2001)\citenamefont{Godefroy, Korb,
  Fleury, and Bryant}}]{Godefroy2001a}
\bibinfo{author}{\bibfnamefont{S.}~\bibnamefont{Godefroy}},
  \bibinfo{author}{\bibfnamefont{J.-P.} \bibnamefont{Korb}},
  \bibinfo{author}{\bibfnamefont{M.}~\bibnamefont{Fleury}}, \bibnamefont{and}
  \bibinfo{author}{\bibfnamefont{R.~G.} \bibnamefont{Bryant}},
  \bibinfo{journal}{Phys. Rev. E} \textbf{\bibinfo{volume}{64}},
  \bibinfo{pages}{021605} (\bibinfo{year}{2001}).

\bibitem[{\citenamefont{Hoop and Prange}(2007)}]{Hoop2007}
\bibinfo{author}{\bibfnamefont{A.~T.~d.} \bibnamefont{Hoop}} \bibnamefont{and}
  \bibinfo{author}{\bibfnamefont{M.~D.} \bibnamefont{Prange}},
  \bibinfo{journal}{J.of Phys. A} \textbf{\bibinfo{volume}{40}},
  \bibinfo{pages}{12463} (\bibinfo{year}{2007}).

\end{thebibliography}

\end{document}